\documentclass[prd,amsfonts,nofootinbib,showpacs,preprint]{revtex4}
\usepackage{hyperref}
\usepackage{graphicx}
\usepackage{dcolumn}
\usepackage{bm}
\usepackage{amsmath}
\usepackage{epsfig}
\usepackage{natbib}
\usepackage{color} 
\usepackage{multirow} 
\usepackage{hhline}

\newcommand{\iu}{{i\mkern1mu}}

\def\21{$\mathrm{SU(2)_L \otimes U(1)_Y}$ }
\def\ew{$\mathrm{SU(2)_L \otimes U(1)_Y}$ }

\newcommand{\AddrAHEP}{AHEP Group, Institut de F\'{i}sica Corpuscular --
  C.S.I.C./Universitat de Val\`{e}ncia, Parc Cientific de Paterna.\\
  C/Catedratico Jos\'e Beltr\'an, 2 E-46980 Paterna (Val\`{e}ncia) - SPAIN}

\newcommand{\Cali}{Universidad Santiago de Cali, Campus
  Pampalinda, Calle 5  No. 6200, 760001, Santiago de Cali, Colombia}

\newcommand{\Cinvestav}{Departamento de F\'{\i}sica, Centro de
  Investigaci{\'o}n y de Estudios Avanzados del IPN\\ Apdo. Postal
  14-740 07000 Mexico, DF, Mexico}
  
\begin{document}

\preprint{IFIC/15-XX}

\title{Updating neutrino magnetic moment constraints}

\author{B. C. Ca\~nas~$^1$}%
\email{bcorduz@fis.cinvestav.mx}
\author{O. G. Miranda~$^1$}%
\email{omr@fis.cinvestav.mx}
\author{A. Parada~$^2$}%
\email{alexander.parada00@usc.edu.co} 
\author{M. T\'ortola~$^3$}%
\email{mariam@ific.uv.es}
\author{J. W. F. Valle~$^3$}
\email{valle@ific.uv.es, URL:
  http://astroparticles.es/} 
\affiliation{$^1$~\Cinvestav}
\affiliation{$^2$~\Cali}
\affiliation{$^3$~\AddrAHEP}
\pacs{13.15.+g,14.60.St,12.60.-i,13.40.Em} 

\begin{abstract}
  In this paper we provide an updated analysis of the neutrino
  magnetic moments (NMMs), discussing both the constraints on the
  magnitudes of the three transition moments $\Lambda_i$ as well as
  the role of the CP violating phases present both in the mixing
  matrix and in the NMM matrix.
  The scattering of solar neutrinos off electrons in Borexino provides
  the most stringent restrictions, due to its robust statistics and
  the low energies observed, below 1 MeV.  Our new limit on the
  effective neutrino magnetic moment which follows from the most
  recent Borexino data is $3.1\times10^{-11}\mu_B$ at 90\% C.L.
  This corresponds to the individual transition magnetic moment
  constraints: $|\Lambda_1| \leq 5.6\times 10^{-11}\mu_B$,
  $|\Lambda_2| \leq 4.0\times 10^{-11}\mu_B$, and $|\Lambda_3| \leq
  3.1\times 10^{-11}\mu_B$ (90\% C.L.), irrespective of any complex
  phase.
  Indeed, the incoherent admixture of neutrino mass eigenstates
  present in the solar flux makes Borexino insensitive to the Majorana
  phases present in the NMM matrix.
  For this reason we also provide a global analysis including  the
  case of reactor and accelerator neutrino sources, and presenting the
  resulting constraints for different values of the relevant CP
  phases.
  Improved reactor and accelerator neutrino experiments will be needed
  in order to underpin the full profile of the neutrino
  electromagnetic properties.

\end{abstract}

\maketitle

\section{Introduction}

Neutrino physics has now reached the precision age characterizing a
mature science. 
Underpinning the origin of neutrino mass remains an open challenge,
whose investigation could help us find our way towards the ultimate
theory of everything~\cite{Valle:2015pba}.
Indeed, the search for new phenomenological signatures associated to
massive neutrinos may yield valuable clues towards the structure of
the electroweak theory beyond the Standard Model (SM).
Although the field is very active, most of the experimental efforts
are devoted to explore the neutrino mass pattern through the study of
oscillations~\cite{Forero:2014bxa,Maltoni:2004ei}. However it is also
of great importance to investigate the implications of dimension-6
non-standard
interactions~\cite{miranda:2004nb,barranco:2005ps,barranco:2007ej} as
well as electromagnetic properties of the
neutrinos~\cite{cisneros:1971nq,Fujikawa:1980yx,Schechter:1981hw,pal:1982rm,kayser:1982br,nieves:1982zt,Shrock:1982sc,akhmedov:1988uk,lim:1987tk}. Here
we focus on the latter, which has also been a lively subject of
phenomenological research in the last few
years~\cite{beacom:1999wx,grimus:2002vb,barranco:2002te,Giunti:2014ixa,Giunti:2015gga,Barranco:2012xj}.
Indeed, different experiments have set constraints coming mainly from
reactor neutrino studies~\cite{li:2002pn,Deniz:2009mu} as well as from
solar neutrino data~\cite{beacom:1999wx,grimus:2002vb}. 
Future tests from experiments measuring coherent neutrino-nucleus
scattering are expected to improve the current bounds on neutrino
electromagnetic
properties~\cite{Wong:2005vg,Wong:2010zzc,Bolozdynya:2012xv,Kosmas:2015sqa,Kosmas:2015vsa}.
Most of the constraints reported by the experiments refer to the case
of a Dirac neutrino magnetic moment, despite the fact that Majorana
neutrinos are better motivated from the theoretical point of
view~\cite{Schechter:1980gr}.  However the Majorana case has been
considered in Refs.~\cite{barranco:2002te,grimus:2002vb} where a more
complete analysis was performed.
Other recent theoretical studies of the neutrino magnetic moment in
the case of Majorana neutrinos can be found
in~\cite{Balantekin:2013sda} and~\cite{Frere:2015pma}.  \\

In this article we perform a combined analysis of reactor, accelerator
and solar neutrino data, in order to obtain constraints on the
Majorana neutrino transition magnetic moments.  We include the most
recent results from the TEXONO reactor experiment~\cite{Deniz:2009mu}, 
as well as the recent results from the Borexino experiment~\cite{Bellini:2011rx}. 
Data from the reactor experiments Krasnoyarsk~\cite{Vidyakin:1992nf},
Rovno~\cite{Derbin:1993wy} and MUNU~\cite{Daraktchieva:2005kn} as well
as the accelerator experiments LAMPF~\cite{Allen:1992qe} and
LSND~\cite{Auerbach:2001wg} are also included.
Moreover, in our analysis we take into account the updated values of
the neutrino mixing parameters as determined in global oscillation
fits~\cite{Forero:2014bxa}, including the value of $\theta_{13}$
implied by Daya-Bay~\cite{An:2012eh,An:2013zwz} and RENO reactor
data~\cite{Ahn:2012nd}, as well as accelerator
data~\cite{Abe:2013hdq}. Moreover, we pay attention to the role of
the, yet unknown, leptonic CP violating phases.

\section{The neutrino magnetic moment}
\label{sect:NMM}

In this section we will establish the notation used in the description
of neutrino magnetic moments. This will be very important in order to
understand the constraints and the differences between Dirac and
Majorana cases.
For the general Majorana case we have the effective
Hamiltonian~\cite{Schechter:1981hw}
\begin{equation}
\textit{H}^{M}_{em} = -\frac{1}{4}\nu^{T}_{L}C^{-1}~ \lambda~\sigma^{\alpha\beta}\nu_{L}F_{\alpha\beta} + h.c. ,
\end{equation}
where $\lambda = \mu - id$ is an antisymmetric complex matrix
$\lambda_{\alpha\beta}=-\lambda_{\beta\alpha}$, so that $\mu^{T}=-\mu$
and $d^{T}=-d$ are imaginary. Hence, three complex or six real
parameters are needed to describe the Majorana neutrino case.\\

On the other hand, for the particular case~\footnote{A Dirac neutrino
  is equivalent to two Majorana neutrinos of same mass and opposite
  CP~\cite{Schechter:1980gr}. Indeed, in two-component form, the three
  Dirac neutrinos are described by a 6$\times 6$ transition moment
  matrix.} of Dirac neutrino magnetic moments, the corresponding
Hamiltonian is given by~\cite{Grimus:2000tq}
\begin{equation}
\textit{H}^{D}_{em} = \frac{1}{2}\bar{\nu}_{R}~\lambda~\sigma^{\alpha\beta}\nu_{L}F_{\alpha\beta} + h.c. ,
\end{equation}
with $\lambda = \mu - id$ being an arbitrary complex matrix.
Hermiticity now implies that $\mu$ and $d$ obey $\mu=\mu^{\dag}$ and
$d=d^{\dag}$.
We should stress that experimental measurements usually constrain some
process-dependent effective parameter combination. Even in the case of
laboratory neutrino experiments, where the initial neutrino flux is
fixed to have a well determined given flavor, there is no sensitivity
to the final neutrino state and therefore several possibilities must
be envisaged.  
For the case of solar neutrino experiments, one needs to take into
account that the original electron neutrino flux experiences
oscillations on its way to the Earth.
Therefore, most of the neutrino magnetic moment constraints discussed
in the literature correspond to restrictions upon some
process-dependent effective parameter. The latter is expressed in
terms of the fundamental parameters describing the transition magnetic
moments and their phases, as well as the neutrino mixing parameters.

From now on we are concerned with the case of three ``genuine'' active
Majorana neutrinos.  As already mentioned, the Dirac case, with three
active plus three sterile neutrinos, would be a particular case of the
six-dimensional Majorana neutrino picture, in which the standard Dirac
magnetic moment is viewed as a transition moment connecting an
``active'' with a ``sterile'' neutrino.

Before we express our results in terms of a general phenomenological
notation, we can illustrate the general features of the neutrino
magnetic moment for the simplest model, namely we consider the case of
Majorana neutrino masses in the standard \ew gauge
theory~\cite{pal:1982rm}, in which case the charged current
contribution gives
\begin{equation}\label{eq:nmm-M}
\mu_{ij} = \frac{3eG_F}{16\pi^2\sqrt{2}}({m_\nu}_i+{m_\nu}_j)
          \sum^\tau_{\alpha=e} i \, {\mathcal Im} \left[ U^*_{\alpha i}U_{\alpha j}\left(\frac{{m_l}_\alpha}{M_{\rm W}}\right)^2 \right] .
\end{equation}
Notice that, in this example, if the masses of the charged leptons
were degenerate, then the off-diagonal transition magnetic moments
would be zero, due to the assumed unitarity of the $U$
matrix. However, in reality, this is not the case and the transition
magnetic moments are nonzero. Moreover, the phases in $\mu_{ij}$ will
be the same as present in the lepton mixing matrix $U$ and, therefore,
could in principle be reconstructed. However, due to the
proportionality with the neutrino mass, the magnetic moments expected
just from the \ew gauge sector are too small to be phenomenologically
relevant.

Although enhanced Majorana transition moments are possible in extended
theories, this discussion is beyond the scope of this paper. However,
we quote, as an illustrative example, the case of an extended model
with a charged scalar singlet $\eta^+$ suggested in
Ref.~\cite{Fukugita:1987ti}. In this case the neutrino transition
magnetic moment would be dominated by a charged Higgs boson
contribution, and has been estimated as
\begin{equation}
\mu_{ij} = e \sum_{k} \frac{f_{ki}g^\dag_{kj}+g_{ik}f^\dag_{kj}}{32\pi^2}
          \frac{{m_l}_k}{{m^2}_\eta}\left(\ln\frac{m^2_\eta}{{m_l}^2_k}-1\right) .
\end{equation}
Indeed, in principle this scalar contribution could be higher than the
one discussed in Eq.~(\ref{eq:nmm-M}). Note that in the case of
Higgs-dominated NMM one could, in principle, introduce new CP phases
in addition to those characterizing the lepton mixing matrix.\\

The above discussion could be translated into a more phenomenological
approach in which the Dirac NMM is described by an arbitrary complex
matrix $ \lambda = \mu + id $ ($\tilde{\lambda}$) in the flavor (or
mass) basis, while for the Majorana case the matrix $ \lambda$ takes
the form
\begin{equation}
\lambda =
\left(\begin{array}{ccc}
0 & \Lambda_{\tau} & -\Lambda_{\mu} \\
-\Lambda_{\tau} & 0 & \Lambda_{e} \\
\Lambda_{\mu} & -\Lambda_{e} & 0 \end{array}\right) , \qquad
\tilde{\lambda} =
\left(\begin{array}{ccc}
0 & \Lambda_{3} & -\Lambda_{2} \\
-\Lambda_{3} & 0 & \Lambda_{1} \\
\Lambda_{2} & -\Lambda_{1} & 0 \end{array}\right) ,
\label{Eq:nmm-matrix}
\end{equation}
where we have used the notation $\lambda_{\alpha\beta} =
\varepsilon_{\alpha\beta\gamma}\Lambda_{\gamma}$, where we assume the
transition magnetic moments $\Lambda_\alpha$ and $\Lambda_i$ to be
complex parameters:
$\Lambda_{\alpha}=|\Lambda_{\alpha}|e^{\iu\zeta_{\alpha}}$,
$\Lambda_{i}=|\Lambda_{i}|e^{\iu\zeta_{i}}$. We now turn to the issue
of extracting information on these parameters from experiment.

\subsection{The effective neutrino magnetic moment}

For the particular case of neutrino scattering off electrons, the
differential cross section for the magnetic moment contribution will
be given by
\begin{equation}\label{eq:xsec_mm}
\left(\frac{d\sigma}{dT}\right)_{{em}}  = \frac{\pi \alpha^{2}}{m^{2}_{e}\mu^{2}_{B}}\left(\frac{1}{T}-\frac{1}{E_{\nu}}\right){\mu_{\nu}}^{2} ,
\end{equation}
where $\mu_{\nu}$ is an effective magnetic moment accounting for the
NMM contribution to the scattering process.\\

The effective magnetic moment $\mu_{\nu}$ is defined in terms of the
components of the NMM matrix in Eq.~(\ref{Eq:nmm-matrix}). In the
flavor basis this can be written as
~\cite{Grimus:2000tq,grimus:2002vb}
\begin{equation}
({\mu_{\nu}^{F}})^{2} = a^{\dag}_{-}\lambda^{\dag}\lambda a_{-} + a^{\dag}_{+}\lambda\lambda^{\dag}a_{+}
\end{equation}
where $a_{-}$\, and \,$a_{+}$ denote the negative and positive helicity
neutrino amplitudes, respectively. One finds
\begin{eqnarray}
({\mu_{\nu}^{F}})^{2} &=& |a^1_{-} \Lambda_\mu -  a^2_{-} \Lambda_e|^2 + 
   |a^1_{-} \Lambda_\tau -  a^3_{-} \Lambda_e|^2 + 
   |a^2_{-} \Lambda_\tau -  a^3_{-} \Lambda_\mu|^2 +  \nonumber \\
  & & |a^1_{+} \Lambda_\mu -  a^2_{+} \Lambda_e|^2 + 
   |a^1_{+} \Lambda_\tau -  a^3_{+} \Lambda_e|^2 + 
   |a^2_{+} \Lambda_\tau -  a^3_{+} \Lambda_\mu|^2 .
   \label{Eq:meff_F}
\end{eqnarray}

In order to write the expression for the effective neutrino magnetic
moment in the mass basis we will need the transformations
\begin{equation}
\tilde{a}_{-} = U^{\dag}a_{-},\qquad \tilde{a}_{+}=U^{T}a_{+},\qquad \tilde{\lambda}=U^{T}\lambda U.
\end{equation}
leading to the expression
\begin{equation}
({\mu_{\nu}^{M}})^{2} = \tilde{a}^{\dag}_{-}\tilde{\lambda}^{\dag}\tilde{\lambda} \tilde{a}_{-} + \tilde{a}^{\dag}_{+}\tilde{\lambda}\tilde{\lambda}^{\dag}\tilde{a}_{+}.
\end{equation}
so that
\begin{eqnarray}
\label{eq:mu-gen}
({\mu_{\nu}^{M}})^{2} &=& |\tilde{a}^1_{-} \Lambda_2 -  \tilde{a}^2_{-} \Lambda_1|^2 + 
   |\tilde{a}^1_{-} \Lambda_3 -  \tilde{a}^3_{-} \Lambda_1|^2 + 
   |\tilde{a}^2_{-} \Lambda_3 -  \tilde{a}^3_{-} \Lambda_2|^2 +  \nonumber \\
  & & |\tilde{a}^1_{+} \Lambda_2 -  \tilde{a}^2_{+} \Lambda_1|^2 + 
   |\tilde{a}^1_{+} \Lambda_3 -  \tilde{a}^3_{+} \Lambda_1|^2 + 
   |\tilde{a}^2_{+} \Lambda_3 -  \tilde{a}^3_{+} \Lambda_2|^2 ,
\end{eqnarray}
where $\tilde{a}^i_{\pm}$ denotes the $i$-th component of the
$\tilde{a}_{\pm}$ vector.  \\

Before starting the calculations of the effective Majorana magnetic
moment parameter combination corresponding to the different
experimental setups we would like to comment on the counting of
relevant complex phases.
First we write the three complex phases in the transition
magnetic moment matrix as $\zeta_1$, $\zeta_2$ and $\zeta_3$.
From the leptonic mixing matrix we have another 3 CP-violating phases:
the Dirac phase characterizing neutrino oscillations, $\delta$, and
the two Majorana phases involved in lepton number violating
processes~\cite{Schechter:1980gr}.
As noticed in Ref.~\cite{Grimus:2000tq}, three of these six complex
phases are irrelevant, as they can be reabsorbed in different ways.  In
what follows we give our results in terms of the Dirac CP phase
$\delta$ and the relative difference between the transition magnetic
moment phases, $\xi_1 = \zeta_3 - \zeta_2$, $\xi_2 = \zeta_3 -
\zeta_1$, $\xi_3 = \zeta_2 - \zeta_1$, of which only two are
independent.

\subsubsection{Effective neutrino magnetic moment at reactor
  experiments.}

We now consider the effective neutrino magnetic moment parameter
relevant for the case of reactor neutrinos. In this case we have an
initial electron antineutrino flux, so that the only non--zero entry
in the flavor basis will be $a^1_{{+}}=1$.  Therefore, from
Eq.~(\ref{Eq:meff_F}) we get the following expression for the
effective Majorana transition magnetic moment strength parameter
describing reactor neutrino experiments:
\begin{equation}\label{mureac_flav}
(\mu^{F}_{R})^{2} = |\Lambda_{\mu}|^{2} + |\Lambda_{\tau}|^{2} .
\end{equation}
which in the mass basis leads to the expression
\begin{eqnarray}\label{eq:mureac}
(\mu^{M}_{R})^{2} &=& {|{\bf \Lambda}|^{2}} - s^{2}_{12}c^{2}_{13}|\Lambda_{2}|^{2} - c^{2}_{12}c^{2}_{13}|\Lambda_{1}|^{2} - s^{2}_{13}|\Lambda_{3}|^{2}\\\nonumber 
&-& 2s_{12}c_{12}c^{2}_{13}|\Lambda_{1}||\Lambda_{2}|\cos\delta_{12}
- 2c_{12}c_{13}s_{13}|\Lambda_{1}||\Lambda_{3}|\cos\delta_{13}\\\nonumber
&-& 2s_{12}c_{13}s_{13}|\Lambda_{2}||\Lambda_{3}|\cos\delta_{23}
\end{eqnarray}
where $c_{ij} = \cos\theta_{ij}$, $s_{ij} = \sin\theta_{ij}$ and
$\delta_{12}= \xi_3$, $\delta_{23}= \xi_2 - \delta$, and $\delta_{13}=
\delta_{12}-\delta_{23}$. As already noted, $\delta$ is the Dirac
phase of the leptonic mixing matrix and $\xi_3 =
\zeta_{2}-\zeta_{1}$,\, $\xi_2 = \zeta_{3}-\zeta_{1}$, are the
relative phases introduced by the presence of the magnetic moment.
This expression takes into account that $\theta_{13}$ is different
from zero, and hence generalizes the previous result given
in~\cite{grimus:2002vb}.\\

It in important to notice that the effective magnetic moment in
Eq.~(\ref{eq:mureac}) implies a degeneracy between the leptonic phase
$\delta$ and those present in the neutrino transition magnetic
moments, $\xi_2$ and $\xi_3$. As a result, it will not be possible to
disentangle these phases without further
independent experimental information. \\

\begin{figure}[t]
\centering
\includegraphics[width=0.7\linewidth]{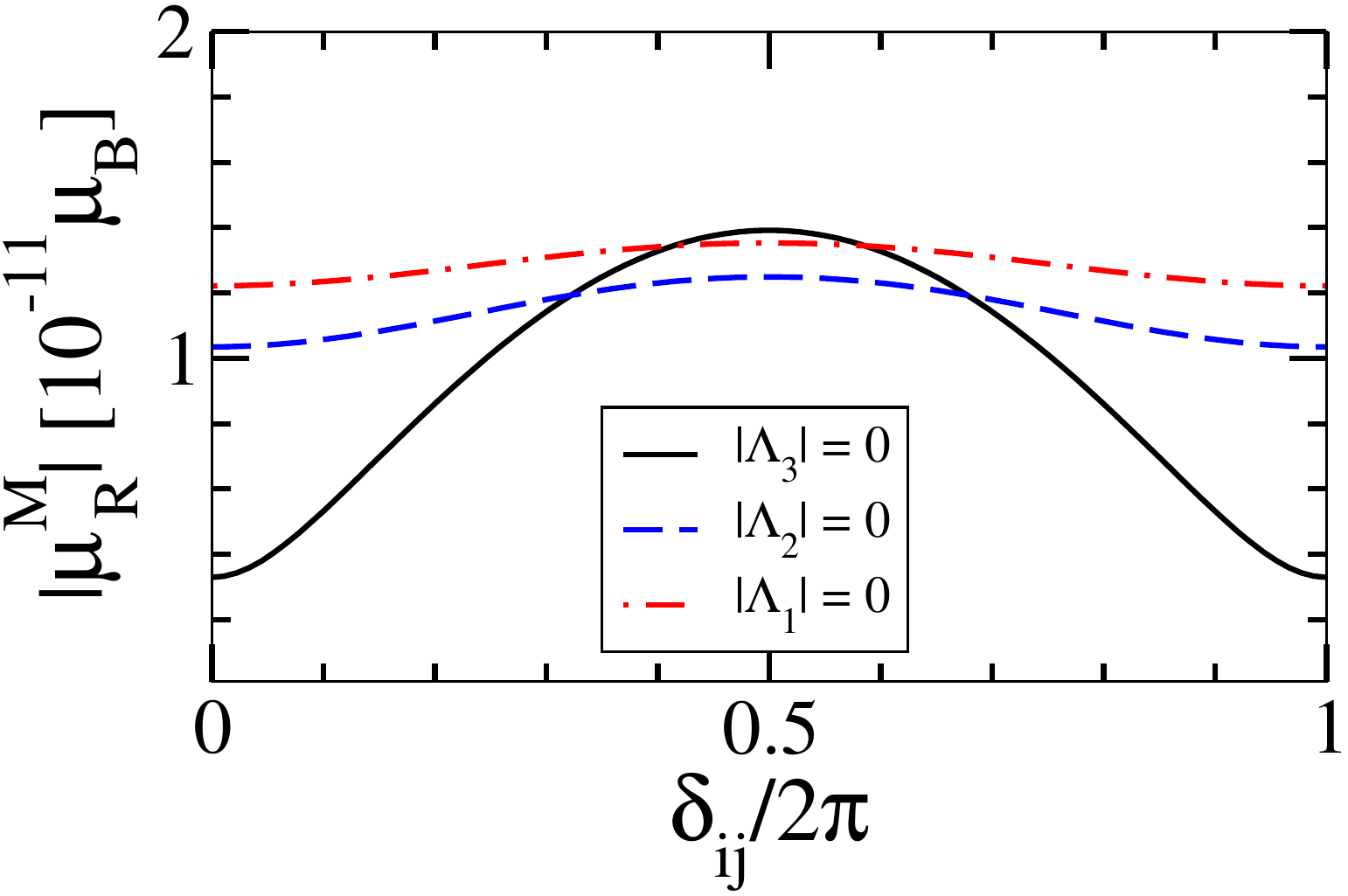}
\caption{Effective Majorana transition magnetic moment probed in
  reactor neutrino experiments, versus the relative phases $\delta_{ij}$
  for three limiting cases where one of the absolute values
  $|\Lambda_k|$ vanishes.  }
\label{fig:reactorphases}
\end{figure}

In order to illustrate the dependence on the different relative phases
$\delta_{ij}$ we show in Fig.~\ref{fig:reactorphases} the value of the
effective Majorana transition magnetic moment for three particular
cases, in which the magnitude of one transition magnetic moment
$|\Lambda_i|$ is assumed to vanish. This implies that the magnetic
moment would depend only on one effective phase $\delta_{ij}$.
Comparing the three curves in Fig.~\ref{fig:reactorphases}, one sees a
strong dependence on the phase $\delta_{12}$ (see solid black line)
while, due to the smallness of $\sin\theta_{13}$, the value of the
phases $\delta_{13}$ and $\delta_{23}$ has little impact on the
magnitude of the effective magnetic moment $\mu_R^M$.

\subsubsection{Effective neutrino magnetic moment at accelerator
  experiments.}

Another relevant measurement for neutrino magnetic moment comes from
accelerator--produced neutrinos arising from pion
decays~\cite{Allen:1992qe,Auerbach:2001wg}. In this case, pion decay
produces a muon neutrino, while the subsequent muon decay generates an
electron neutrino plus a muon antineutrino.  We can write the
effective magnetic moment strength parameter in the flavor basis,
considering for the moment the same proportion of $\nu_e$, $\nu_\mu$
and $\bar{\nu}_\mu$ ($a^1_{-}=1$, $a^2_{-}=1$, $a^2_{+}=1$):
\begin{equation}\label{eq:muAcc}
(\mu^{F}_{A})^{2} = |\mathbf{\Lambda}|^{2} + |\Lambda_{e}|^{2} + 2\,|\Lambda_{\tau}|^{2} - 2\,|\Lambda_{\mu}||\Lambda_{e}|\cos\eta ,
\end{equation}
where $|\mathbf{\Lambda}|^{2} = |\Lambda_{e}|^{2} +
|\Lambda_{\mu}|^{2} + |\Lambda_{\tau}|^{2}$ and $\eta =
\zeta_{e}-\zeta_{\mu}$ is the relative phase between the transition
magnetic moments $\Lambda_e$ and $\Lambda_\mu$.\\

The corresponding expression for the effective neutrino magnetic
moment strength parameter in the mass basis, for $\theta_{13}=0$ will
be given by
\begin{eqnarray}\label{eq:muAcc1}
(\mu^{M}_{A})^{2} & = & |\Lambda_{1}|^{2}[2-(c^{2}_{23}-s^{2}_{23})s^{2}_{12}+2s_{12}c_{12}c_{23}]\\\nonumber
&+& |\Lambda_{2}|^{2}[2-(c^{2}_{23}-s^{2}_{23})c^{2}_{12}-2s_{12}c_{12}c_{23}] + |\Lambda_{3}|^{2}[1+2c^{2}_{23}]\\\nonumber
&+& 2|\Lambda_{1}||\Lambda_{2}|\cos\xi_3[s_{12}c_{12}(c^{2}_{23}-s^{2}_{23})-(c^{2}_{12}-s^{2}_{12})c_{23}] \\\nonumber
&+& 2|\Lambda_{1}||\Lambda_{3}|\cos\xi_2[-c_{12}s_{23}+2s_{12}s_{23}c_{23}]\\\nonumber
&+& 2|\Lambda_{2}||\Lambda_{3}|\cos(\xi_{3}-\xi_2)[-s_{12}s_{23}-2c_{12}s_{23}c_{23}]
\end{eqnarray}
As expected, the Dirac CP phase $\delta$ present in oscillations does
not enter in this expression, and therefore only the two Majorana
phases from the NMM matrix $\xi_2$ and $\xi_3$ are present.
Note however that in our numerical analysis we have used the full
expression with $\theta_{13}\not= 0$, as

\begin{eqnarray}\label{eq:muAcc2}
(\mu^{M}_{A})^{2} & = &  |\Lambda_{1}|^{2} \left[ \sin 2\theta_{12} c_{13}c_{23} + c^{2}_{12}(2c^{2}_{23}+\sin 2\theta_{13} s_{23}\cos\delta)     \right.    \\\nonumber
&+& \left. c^{2}_{13}(s^{2}_{12}+2s^{2}_{23})+s_{13}(s_{13}+2s^{2}_{12}s_{13}s^{2}_{23}-\sin 2\theta_{12}\sin 2\theta_{23}\cos\delta)\right] \\\nonumber
&+& \frac{1}{4}|\Lambda_{2}|^{2}\left[ 8-\cos 2\theta_{23}(1+3\cos 2\theta_{12} +2\cos 2\theta_{13} s^{2}_{12})+4s^{2}_{12}\sin 2\theta_{13} s_{23}\cos\delta \right. \\\nonumber
&+& \left. 4\sin 2\theta_{12} (-c_{13}c_{23}+s_{13}\sin 2\theta_{23}\cos\delta)\right]+|\Lambda_{3}|^{2}\left(2+c^{2}_{13}\cos 2\theta_{23}-\sin 2\theta_{13} s_{23}\cos\delta\right)\\\nonumber
&+& 2|\Lambda_{1}||\Lambda_{2}|\left\{\cos\xi_{3}\left[-c^{2}_{12}c_{13}c_{23}+c_{23}(s^{2}_{12}c_{13}+\sin 2\theta_{12} c_{23}) \right.\right.\\\nonumber 
&+& \left. s_{12}c_{12}(-1+\cos 2\theta_{23} s^{2}_{13}+\sin 2\theta_{13} s_{23}\cos\delta)\right]\\\nonumber 
&+&\left. s_{13}\sin 2\theta_{23}(\cos 2\theta_{12}\cos\delta\cos\xi_{3}+\sin\delta\sin\xi_{3})\right\}\\\nonumber
&+& |\Lambda_{1}||\Lambda_{3}|\left\{ 2\cos(\xi_{2}-\delta)(-c_{12}c_{13}\cos 2\theta_{23}+s_{12}c_{23})s_{13} \right.\\\nonumber
&+&2 \left.\left[c_{13}\cos\xi_{2}(-c_{12}c_{13}+2s_{12}c_{23})
+ c_{12}s^{2}_{13}\cos(\xi_{2}-2\delta)\right]s_{23}\right\}\\\nonumber
&-& 2|\Lambda_{2}||\Lambda_{3}|\left\{\frac{1}{2}s_{12}\cos(\xi_{1}-\delta)(\cos 2\theta_{23}\sin 2\theta_{13}+2\cos 2\theta_{13} s_{23}\cos\delta) \right. \\\nonumber
&+& \left. c_{12}\left[c_{23}s_{13}\cos(\xi_{1}-\delta)+c_{13}\sin 2\theta_{23} \cos\xi_{1}\right]+s_{12}s_{23}\sin\delta\sin(\delta-\xi_{1})\vphantom{\frac{1}{2}}\right\}\\\nonumber
\end{eqnarray}
Notice that we have used here the phase $\xi_1 =
\xi_2-\xi_3$. Although this is not an independent phase, it is hepful
to simplify the previous formula. Therefore, the final expression is
given in terms of the three independent phases $\delta$, $\xi_2$ and
$\xi_3$. One can check that in the limit $\theta_{13}$ = 0, the
expression in Eq.~(\ref{eq:muAcc1}) is recovered.

\subsubsection{Effective neutrino magnetic moment in Borexino.}

Here we calculate the effective magnetic moment strength parameter
relevant for experiments measuring solar neutrinos through their
scattering with electrons, like Borexino~\footnote{The same result
  will apply for the Super-Kamiokande experiment, not included here
  due to its smaller sensitivity to the neutrino magnetic
  moment~\cite{grimus:2002vb}.}.
In this case, the electron neutrinos originally produced in the solar
interior undergo flavor oscillation and they arrive to the Earth
detector as an incoherent sum of mass eigenstates.  Using the
well-justified approximation where~\cite{grimus:2002vb}
\begin{equation}
P^{3\nu}_{e3}  = \sin^2\theta_{13}, \quad 
P^{3\nu}_{e1}  = \cos^2\theta_{13}P^{2\nu}_{e1}, \quad 
P^{3\nu}_{e2}  = \cos^2\theta_{13}P^{2\nu}_{e2}, \quad 
\end{equation}
with $P^{2\nu}_{ej}$ ($j=1,2$) being the effective two-neutrino
oscillation probabilities for solar neutrinos, we arrive to the
effective neutrino magnetic moment strength parameter in the mass
basis,
\begin{equation}\label{eq:nmm_sun}
(\mu^{M}_{\rm{sol}})^{2} = |\mathbf{\Lambda}|^{2} -
  c^{2}_{13}|\Lambda_{2}|^{2} + (c^{2}_{13}-1)|\Lambda_{3}|^{2} +
  c^{2}_{13}P^{2\nu}_{e1}(|\Lambda_{2}|^{2}-|\Lambda_{1}|^{2}).
\end{equation}
where the unitarity condition, $P^{2\nu}_{e1}+P^{2\nu}_{e2} = 1$, has
also been assumed.  The calculation of this expression in the flavor
basis is more complicated due to presence of the neutrino transition
probabilities and therefore we do not include it here.

As we can see from Eq.~(\ref{eq:nmm_sun}), the expression of the
effective magnetic moment for solar neutrinos is independent of any
phase, as has already been noticed~\cite{grimus:2002vb}. Here we take
into account the non-zero value of $\theta_{13}$ for the first time in
this kind of analysis. Taking advantage of the previous equation we
obtain constraints on the individual neutrino transition magnetic
moments.
After obtaining the neutrino magnetic moment expressions for the case
of $\theta_{13}\neq 0$, we now turn our attention to the relevant
experiments for our analysis.

\section{Neutrino data analysis}

Having evaluated the effective neutrino magnetic moment strength
parameter for reactor, accelerator and solar neutrino experiments, we
are ready to perform a combined analysis of the experimental data in
order to get constraints on the three different transition magnetic
moments $\Lambda_i$.  In order to perform this analysis we make some
assumptions on the phases $\delta$, $\xi_2$ and $\xi_3$.  In the next
section we will describe the data used in the fit and show the
results.  We now briefly describe the statistical analysis performed
in this article.

\subsection{Reactor antineutrinos}
\label{subsec:reac}

We start by describing the reactor antineutrino experiments. They use
the antineutrino flux coming from a nuclear reactor, in combination
with a detector sensitive to the electron antineutrino scattering off
electrons. The total number of events (in the $i$-th bin) in these
experiments is given by
\begin{equation}
N_{R}^i=\kappa\int dE_{\nu}\int dT\int^{T_{i+1}}_{T_{i}}dT'\lambda(E_{\nu})\frac{d\sigma}{dT}(E_{\nu},T,{\mu})R(T,T') ,
\end{equation}
where the integrals run over the detected electron recoil energy $T'$,
the real recoil energy $T$, and the neutrino energy $E_\nu$.
$T_i$ and $T_{i+1}$ are the minimum and maximum energy of the $i$-th
bin, respectively.
The parameter $\kappa$ stands for the product of the total number of
targets times the total antineutrino flux times the total exposure
time of the experimental run and $\lambda(E_\nu)$ is the antineutrino
energy spectrum coming from the nuclear
reactor~\cite{Mueller:2011nm,Kopeikin:1997ve}. Some of the experiments
under consideration reported their resolution function $R(T,T')$,
given by
\begin{equation}
R(T,T') = \frac{1}{\sqrt{4\pi}\sigma}\exp{\left(\frac{-(T-T')^{2}}{2\sigma^{2}}\right)} .
\end{equation}
where $\sigma$ stands for the error in the kinetic energy
determination.  When the information on this resolution function is
not available, we have assumed perfect energy resolution and taken it
as a delta function: $R(T,T') = \delta(T-T')$.

Finally, the standard differential cross section for the process of
$\bar{\nu}_e$-electron scattering is given by
\begin{equation}
\frac{d\sigma}{dT} = \frac{2G^2_Fm_e}{\pi}
              \left[g^2_R + g^2_L (1-\frac{T}{E_\nu})^2 - g_Lg_R m_e\frac{T}{E_\nu^2}\right] ,
\end{equation}
where $m_e$ is the electron mass and $G_F$ is the Fermi constant.  For
this process, at tree level, the coupling constants $g_{L,R}$ are
given by $g_L = 1/2 + \sin^2\theta_{\rm W}$ and $g_R =
\sin^2\theta_{\rm W}$.
The assumed non-zero neutrino magnetic moment yields a new
contribution to the cross section, given by
\begin{equation}
\left(\frac{d\sigma}{dT}\right)_{{em}}  = \frac{\pi \alpha^{2}}{m^{2}_{e}}
\left(\frac{1}{T}-\frac{1}{E_{\nu}}\right){\mu_{R}}^{2}\, ,
\end{equation}
where $\mu_{R}=\mu^{F,M}_{R}$ is the reactor effective neutrino
magnetic moment, either in the mass or flavor basis, as already
discussed in Eqs.~(\ref{mureac_flav}) and (\ref{eq:mureac}).
This gives rise to an additional neutrino signal at reactor
experiments.
\begin{table}
\caption{90\% C.L. limits (95\% C.L. for Rovno) on the effective
  neutrino magnetic moment from reactor and accelerator data.}
\begin{center}
\begin{tabular}{l c}
\hline
 Experiment &  Bounds \\
\hline
  Reactors &   [Expression in Eqs.(\ref{mureac_flav})-(\ref{eq:mureac})]  \\
\hline
KRASNOYARSK~\cite{Vidyakin:1992nf} 
                  	                   &  $\mu_{\bar{\nu}_e}\leq$ $ 2.7 \times 10^{-10}\mu_{B}$ \\
ROVNO~\cite{Derbin:1993wy}  
                                     &  $\mu_{\bar{\nu}_e}\leq 1.9 \times 10^{-10}\mu_{B}$ \\
MUNU~\cite{Daraktchieva:2005kn}
                                      &  $\mu_{\bar{\nu}_e}\leq 1.2 \times 10^{-10}\mu_{B}$ \\
TEXONO~\cite{Deniz:2009mu}
                                     &  $\mu_{\bar{\nu}_e}\leq 2.0 \times 10^{-10}\mu_{B}$ \\
\hline
 Accelerators &  [Expression in Eqs.~(\ref{eq:muAcc})-(\ref{eq:muAcc1})-(\ref{eq:muAcc2})] \\
\hline
LAMPF~\cite{Allen:1992qe}  
                           & $\mu_{\nu_e}\leq 7.3 \times 10^{-10}\mu_{B}$ \\
LAMPF~\cite{Allen:1992qe}  
                           
                            & $\mu_{\nu_\mu}\leq 5.1 \times 10^{-10}\mu_{B}$ \\
LSND~\cite{Auerbach:2001wg} 
                            & $\mu_{\nu_e}\leq 1.0 \times 10^{-9}\mu_{B}$ \\
LSND~\cite{Auerbach:2001wg} 
                            & $\mu_{\nu_\mu}\leq 6.5 \times 10^{-10}\mu_{B}$ \\
\hline
\end{tabular}
\label{tab:reac-acc}
\end{center}
\end{table}
Finally, we perform our statistical analysis using the following
$\chi^2$ function:
\begin{equation}\label{eq:chi2}
\chi^{2}=\sum^{N_{bin}}_{i=1}
\left(\frac{O^i_R- N_R^i(\mu_R)} {\Delta_i}\right)^{2},
\end{equation}
where $O_R^i$ and $N_R^i$ are the observed number of events and the
predicted number of events in the presence of an effective magnetic
moment $\mu_R$ at the $i$-th bin, respectively.  Here $\Delta_i$ is the
statistical error at each bin.

In our analysis, we have used the experimental results
reported by Krasnoyarsk~\cite{Vidyakin:1992nf},
Rovno~\cite{Derbin:1993wy}, MUNU~\cite{Daraktchieva:2005kn}, and
TEXONO~\cite{Deniz:2009mu} reactor experiments.  
As a first step we have calibrated our numerical analysis by
reproducing the constraints on the effective neutrino magnetic moment
reported by each experiment.  To do this we performed an analysis as
similar as possible to the original references, using the antineutrino
spectrum description available at the time of the corresponding
experiment as well as the antineutrino electron cross section.
Afterwards, we have recalculated our limits on the NMM by introducing
the new antineutrino reactor spectrum.
Our results on reactor neutrino experiments are summarized in 
  the upper part of Table~\ref{tab:reac-acc}.

Although it is not listed in Table~\ref{tab:reac-acc}, we have also
analyzed the case of the GEMMA~\cite{Beda:2012zz} experiment. In this
case there is no detection of the SM signal and therefore, the
statistical analysis is a bit different from what we have described
above.  It is important to notice that this experiment gives a
stronger constraint compared with other reactor experiments
($\mu_{\bar{\nu}_e}\leq 2.9 \times 10^{-11}\mu_{B}$). However, the
different statistical treatment employed to analyze GEMMA's data makes
it difficult to establish a direct comparison with the remaining
reactor results.

\subsection{Accelerator data}

For the case of accelerator neutrinos we have considered the
experimental data reported by the LAMPF~\cite{Allen:1992qe} and
LSND~\cite{Auerbach:2001wg} collaborations.  The expected number of
events for electron and muon neutrinos is calculated as
\begin{equation}
N_{A}=\int dE_{\nu}\int^{T_{f}}_{T_{i}}dT'\lambda(E_{\nu})\frac{d\sigma}{dT}(E_{\nu},T',\mu),
\end{equation}
where $A$ refers to the type of event ($\nu_e$, $\nu_\mu$ or
$\bar{\nu}_\mu$), $E_{\nu}$ corresponds to the neutrino energy, $T'$
is the electron recoil energy, and $\lambda(E_{\nu})$ is the neutrino
energy spectrum from the accelerator
experiments~\cite{Allen:1992qe,Auerbach:2001wg}.
The statistical analysis is similar to the one for reactor
antineutrinos described in the previous subsection.
As a first step we try to reproduce the individual limits on the
magnetic moment of electron and muon neutrinos reported by the
experimental collaborations.  To do this we have used the $\chi^{2}$
function given by Eq.~(\ref{eq:chi2}), comparing the expected event
number reported by the experiments with the calculated number of
events.  The limits on the muon and electron neutrino magnetic moments
are derived taking into account the following relations for the
effective neutrino magnetic moment (see Refs.~\cite{Allen:1992qe}
and~\cite{Auerbach:2001wg} for details): $\mu^{2}_{\nu_e} +
\alpha\mu^{2}_{\nu_\mu} < \mu^{2}_{eff}$, where $\alpha$ stands for
the rate between muon and electron neutrinos in the detector.  This
ratio is expected to be equal to two as first approximation, since
each pion decay produces a muon antineutrino plus a muon neutrino plus
an electron neutrino. The values reported by the experimental
collaborations are $\alpha = 2.1$ for LAMPF~\cite{Allen:1992qe} and
$\alpha = 2.4$ for LSND~\cite{Auerbach:2001wg}.
The limits on the effective neutrino magnetic moment derived from
LAMPF and LSND data are reported in the lower part of
Table~\ref{tab:reac-acc}.  For the more complete analysis including
the complex phases in the neutrino magnetic moment matrix we take
$\alpha=2$, as included in Eqs.~(\ref{eq:muAcc})-(\ref{eq:muAcc2}).

\subsection{Borexino data}

The Borexino experiment has successfully measured a large part of the
neutrino flux spectrum coming from the
Sun~\cite{Arpesella:2007xf,Bellini:2008mr,Bellini:2013lnn,Bellini:2014uqa}
and has set limits on the effective neutrino magnetic moment by using
their observations of the Beryllium solar neutrino
line~\cite{Back:2002cd,Arpesella:2008mt}. In this paper we will
consider the more recent measurements of the Beryllium solar flux
reported in Ref.~\cite{Bellini:2011rx} in order to obtain a stronger
constraint.

For reactor and accelerator experiments, our statistical analysis
followed the covariant approach. In the case of the Borexino, however,
we have adopted the pull approach~\cite{Fogli:2002pt}.
Focusing on the Beryllium neutrino flux, the expected number of events
at the $i$-th bin, $N_{i}^{th}$, will be given by
\begin{equation}\label{NTH_pull}
N_i^{th}=\kappa \int
\frac{d\sigma}{dT_{e}}(E_{\nu},T_{e})R(T_{e},T'_{e})dT_{e}dT'_{e}+N_{i}^{bg},
\end{equation}
where $N_{i}^{bg}$ represents the number of expected background events
at the considered energy bin.  Here $\kappa$ stands for the product of
the number of target electrons, the detection time window (740.7 days
in this case), and the total Beryllium neutrino flux. $T_{e}$ is the
real electron kinetic energy and $T'_{e}$ is the reconstructed one.
The energy resolution function $R(T_{e},T'_{e})$ of the experiment is
given by
\begin{equation}
R(T_{e},T'_{e})=
\frac{1}{\sqrt{2\pi} \sigma^2 }exp\left(\frac{(T_{e}-T'_{e})^2}{2\sigma^2}\right)
\end{equation}
with $\sigma/T_{e}=0.06\sqrt{T_{e}/\rm{MeV}}$~\cite{GonzalezGarcia:2009ya}.
Finally the differential cross section is given by 
\begin{equation}
\frac{d\sigma_{\alpha}}{dT_{e}}(E_{\nu}, T_{e}) = \overline{P}_{ee}\frac{d\sigma_{e}}{dT_{e}}(E_{\nu},T_{e}) + (1-\overline{P}_{ee})\frac{d\sigma_{\mu-\tau}}{dT_{e}}(E_{\nu},T_{e}),
\label{eq:diff-cross}
\end{equation}
where the average survival electron-neutrino probability for the Beryllium line,  
$\overline{P}_{ee}$, determines the flavour composition of the neutrino flux detected in the experiment.
According to the most recent analysis of solar neutrino data in
Ref.~\cite{Forero:2014bxa} (excluding Borexino data to avoid any
correlation with the present analysis) this value is set to
$\overline{P_{ee}^{\rm{th}}} = 0.54\pm 0.03$. 

\begin{table}
  \caption{90\% C.L. limits on the effective neutrino magnetic moment from  
    Borexino data. We show for comparison the constraint previously reported and the 
    bound obtained in this work}
\begin{center}
\begin{tabular}{l p{5mm} c p{5mm} c  c}
\hline Experiment &  &Previous limit~\cite{Arpesella:2008mt} 
                        &  &This work &  Full expression\\ 
\hline 
Borexino &  &$\mu_{\nu}\leq 5 \times 10^{-11}\mu_{B}$  
         & &$\mu_{\nu}\leq 3.1 \times 10^{-11}\mu_{B}$  & Eq.~(\ref{eq:nmm_sun})\\ 
\hline
\end{tabular}
\label{table:BorexinoMM}
\end{center}
\end{table}

In order to explore the sensitivity of the Borexino experiment to the
neutrino magnetic moments, we include the new contribution to the
differential cross section in Eq.~(\ref{eq:diff-cross}):
\begin{equation}\label{eq:xsec_borexino}
\left(\frac{d\sigma}{dT}\right)_{{em}}  = \frac{\pi \alpha^{2}}{m^{2}_{e}\mu^{2}_{B}}\left(\frac{1}{T}-\frac{1}{E_{\nu}}\right){\mu_\mathrm{sol}}^{2} ,
\end{equation}
where $\mu_\mathrm{sol}$ is the effective magnetic moment strength
parameter relevant for the Borexino solar neutrino experiment derived
in Eq.{~(\ref{eq:nmm_sun}) in the mass basis.
  This yields a new contribution to the expected number of events,
  which will determine the sensitivity to the presence of a neutrino
  magnetic moment.

  With the expected event number, we have fitted our predictions to
  the experimental data in the statistical analysis. There we have
  considered the Borexino systematic errors associated to the fiducial
  mass ratio uncertainty ($\pi_{vol}=6\%$), the energy scale
  uncertainty ($\pi_{scl}^b=1\%$) and the energy resolution
  uncertainty ($\pi_{res}=10\%$).  We have also included in the fit
  the electron-neutrino survival probability $\overline{P}_{ee}$ as a
  free parameter (using the value of $\overline{P_{ee}^{\rm{th}}}$
  given above as a prior) with the corresponding penalty in the
  $\chi^2$ function.
  The constraint we have obtained for the effective neutrino magnetic
  moment using the latest Borexino data is given in
  Table~\ref{table:BorexinoMM}. For comparison, we have also included
  in the table the previous bound, derived by the Borexino
  Collaboration in Ref.~\cite{Arpesella:2008mt}. Note that our updated
  limit is comparable to the strongest bound reported by the GEMMA
  experiment and previously discussed in Sect.~\ref{subsec:reac}.

  Using the expression of the effective neutrino magnetic moment in
  Borexino given by Eq.~(\ref{eq:nmm_sun}), we can also obtain limits
  on the individual elements of the transition magnetic moment matrix
  $\Lambda_i$.  In this case, the calculations involve the neutrino
  oscillation probability $P^{2\nu}_{e1}$, which, as before, is
  considered in our $\chi^2$ analysis as a free parameter with an
  associated penalty term. As a prior, we have considered again the
  value of the probability predicted by the analysis of all other
  solar neutrino data except Borexino, given by
  $P^{2\nu}_{e1}\big|_{\rm th} =0.61\pm 0.06$~\cite{Forero:2014bxa}.
Our results are summarized in the last row of Table \ref{table:NMM-one}.

\section{Limits on the neutrino magnetic moment}

In the previous section we have derived bounds on the effective
neutrino magnetic moment parameter combinations relevant in reactor,
accelerator and solar neutrino experiments. Our results are summarized
in Tables \ref{tab:reac-acc} and \ref{table:BorexinoMM}.  The most
remarkable result is the limit obtained with the latest Borexino data:
$\mu_\mathrm{sol} \leq 3.1 \times 10^{-11}\mu_{B}$ , which is
comparable to the constraint reported by the GEMMA~\cite{Beda:2012zz}
collaboration using reactor antineutrinos, $\mu_\mathrm{R} \leq 2.9
\times 10^{-11}\mu_{B}$~\footnote{Both limits correspond to 90\% C.L.}.\\

One can go one step further and make a combined analysis using all the
data studied so far.
This combined study can not be done in terms of the effective magnetic
moments, since they are different for each type of experiment, but we
need to use a more general formalism, as the one we have discussed in
section~\ref{sect:NMM}.
We choose to work in the mass basis and hence we consider the NMM
parameters $\Lambda_1$, $\Lambda_2$ and $\Lambda_3$.
As a first step in our analysis, we take all elements as real, setting
the complex phases to zero, and we also take one nonzero transition
magnetic moment element $\Lambda_i$ at a time. The results from this
analysis are shown in Table~\ref{table:NMM-one}, where one sees that
the Borexino constraint is considerably stronger than the others,
except for GEMMA, as we already commented~\footnote{Due to the
  complexity of the statistical analysis in GEMMA, here we have only
  translated their reported bound~\cite{Beda:2012zz} into $\Lambda_i$
  by using Eq.~(\ref{eq:mureac}), instead of including GEMMA data
  explicitly in the global analysis.}.
\begin{table}[!t]
  \caption{    90\% C.L. limits  on the neutrino magnetic moment components in the
    mass basis, $\Lambda_i$, from reactor, accelerator, and solar data
    from Borexino. In this particular analysis we constrain one parameter at 
    a time, setting all other magnetic moment parameters and phases to
    zero.\\}
\begin{tabular}{l p{12mm} c p{1cm} c p{1cm} c}
\hline
Experiment & & $|\Lambda_{1}|$ &&  $|\Lambda_{2}|$ & & $|\Lambda_{3}|$\\ [0.5ex]
\hline
KRASNOYARSK & & $ 4.7\times10^{-10}\mu_{B}$& & $3.3\times10^{-10}\mu_{B}$& & $2.8\times10^{-10}\mu_{B}$\\
ROVNO & & $ 3.0 \times 10^{-10}\mu_{B}$&& $2.1\times10^{-10}\mu_{B}$& & $1.8\times10^{-10}\mu_{B}$ \\
MUNU & & $2.1 \times 10^{-10}\mu_{B}$&& $1.5\times10^{-10}\mu_{B}$& & $1.3\times10^{-10}\mu_{B}$ \\
TEXONO & & $ 3.4 \times 10^{-10}\mu_{B}$ & &$2.4\times10^{-10}\mu_{B}$& & $2.0\times10^{-10}\mu_{B}$ \\
GEMMA & & $ 5.0 \times 10^{-11}\mu_{B}$ & &$3.5\times10^{-11}\mu_{B}$& & $2.9\times10^{-11}\mu_{B}$ \\
\hline
LSND & & $ 6.0 \times 10^{-10}\mu_{B}$ & & $8.1\times10^{-10}\mu_{B}$ & & $7.0\times10^{-10}\mu_{B}$ \\
LAMPF & & $ 4.5\times10^{-10}\mu_{B}$ & & $ 6.2\times10^{-10}\mu_{B}$ & & $5.3\times10^{-10}\mu_{B}$  \\
\hline
Borexino   & &$5.6\times 10^{-11}\mu_{B}$ & & $ 4.0\times 10^{-11}\mu_{B}$ & & $3.1\times 10^{-11}\mu_{B}$ \\ 
\hline
\end{tabular}
\label{table:NMM-one}
\end{table}

We have also considered a more complete analysis taking into account
the role of the phases in the reactor and accelerator data. Notice
that the effective magnetic moment for the Borexino experiment is
independent of all the complex phases (see Eq.~(\ref{eq:nmm_sun}))
since solar neutrinos arrive to the Earth as an incoherent sum of mass
eigenstates and therefore, no interference terms appear in the
calculation.
For the case of reactor neutrinos, we have performed a statistical
analysis of TEXONO data~\cite{Deniz:2009mu} for different choices of
the complex phases of $\Lambda_i$, $\zeta_i$, and taking all
transition magnetic moment amplitudes as nonzero. The results of this
analysis are shown in Fig.~\ref{fig.reactor.phase}. There we present
the 90\% C.L. allowed regions for the transition magnetic moments in
the mass basis in the form of two-dimensional projections in the
planes ($|\Lambda_i|$, $|\Lambda_j|$).  In all cases the regions have
been obtained marginalizing over the undisplayed parameter
$|\Lambda_k|$.  Concerning the complex phases, in the two cases
considered we have fixed the mixing matrix CP phase $\delta$ to its
currently preferred value~\cite{Forero:2014bxa}: $\delta=3\pi /2$. For
the complex phases in the transition magnetic moments we have
considered two cases.  The magenta (outer) region in
Fig.~\ref{fig.reactor.phase} corresponds to the case with all phases
equal to zero: $\xi_2 = \xi_3 = 0$ while the orange (inner) displayed
region has been obtained for $\xi_2=0$ and $\xi_3 =\pi/2$.
\begin{figure}[t]
\centering
\includegraphics[width=0.75\linewidth]{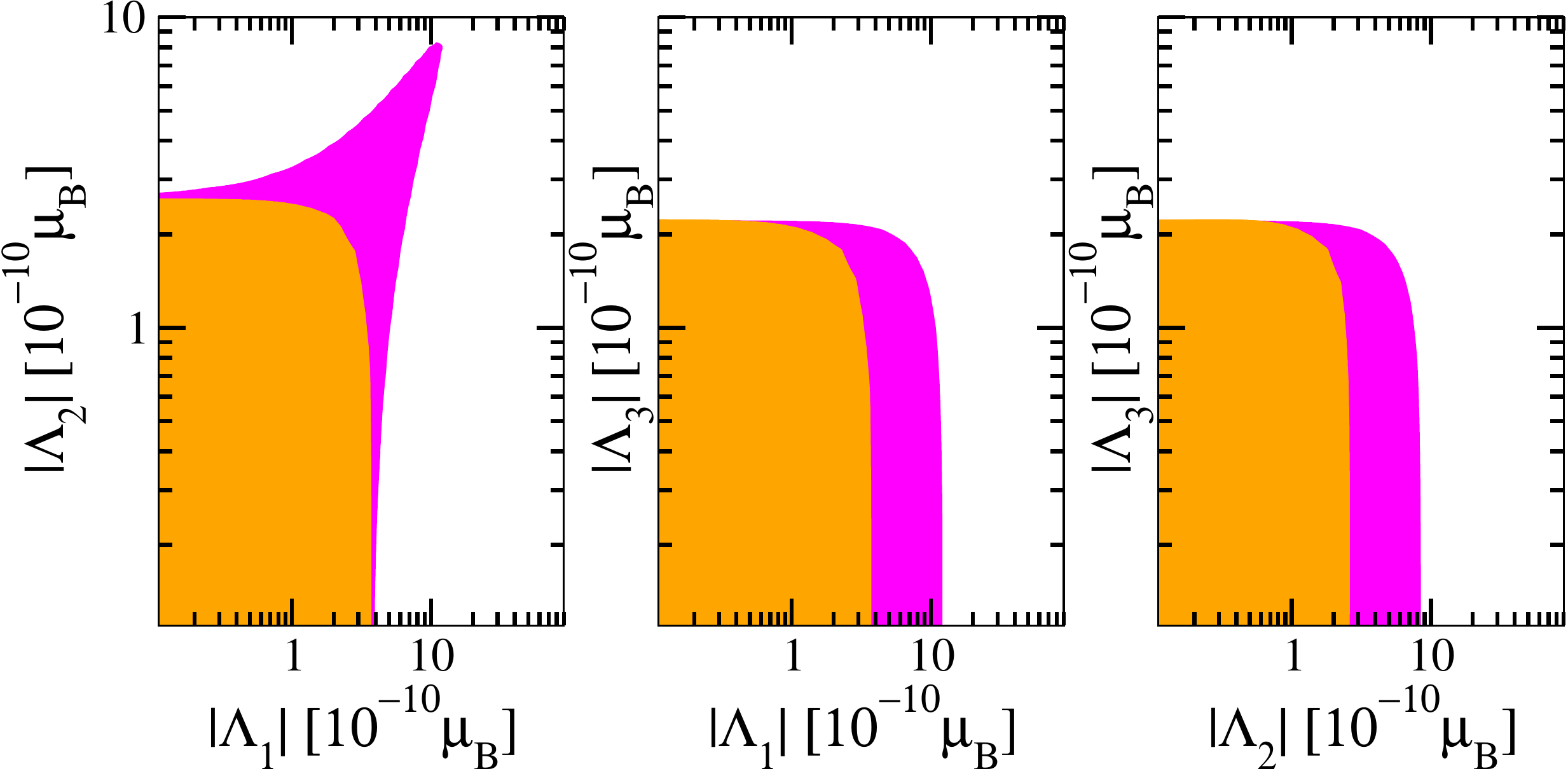}
\caption{90\% C.L. allowed regions for the transition neutrino magnetic moments in 
 the mass basis from the reactor experiment TEXONO. 
  The two-dimensional projections in the plane ($|\Lambda_i|$, $|\Lambda_j|$) 
  have been calculated marginalizing over the third component.
  The magenta (outer) region  is obtained for $\delta = 3\pi/2$ and $\xi_2 = \xi_3 = 0$, 
  while the orange (inner) region appears for 
  $\delta=3\pi/2$, $\xi_2=0$ and $\xi_3 =\pi/2$. %
}
\label{fig.reactor.phase}
\end{figure}
One can see in this plot the role of the CP phases, since the
resulting restrictions on the transition magnetic moments
$|\Lambda_1|$ and $|\Lambda_2|$ depend on the chosen phase
combinations.
Note, however, that in the two cases analyzed the bound on
$|\Lambda_3|$ is practically unchanged, showing that in this
particular case the complex phases are not very relevant. As discussed
in Fig.~\ref{fig:reactorphases}, this is due to the fact that the
terms involving simultaneously $|\Lambda_3|$ and any complex phase in
the expression of the effective magnetic moment in
Eq.~(\ref{eq:mureac}) are proportional the small quantity
$\sin\theta_{13}$ and therefore they are subdominant with respect to
the real terms in $\mu_R^M$.

Finally, we have performed a combined analysis of all the reactor and
accelerator data discussed in this paper, for a particular choice of
phases ($\delta=3\pi / 2$ and $\xi_i = 0$) and compared it with the
corresponding $\chi^2$ analysis of Borexino data. The results, shown
in Fig.~\ref{fig:combined}, illustrate how Borexino is more
sensitive in constraining the magnitude of the transition neutrino
magnetic moments. Since the Borexino effective magnetic moment depends
only on the square magnitudes of these transition magnetic moments,
its constraints are in practice the same as those in the
one-parameter-at-a-time analysis. In this sense, a detailed analysis
of GEMMA data, not performed here, is not expected to give a result as
robust as the one obtained with Borexino data. However, one should
notice that future reactor and accelerator experiments are the only
ones that could give information on individual transition magnetic
moments as well as on the Majorana phases discussed here, an
information inaccessible at Borexino.
This information is crucial in certain analyses of the neutrino
Majorana nature such as those recently performed in
Refs.~\cite{Balantekin:2013sda,Frere:2015pma}.

\begin{figure}[t]
\centering
\includegraphics[width=0.75\linewidth]{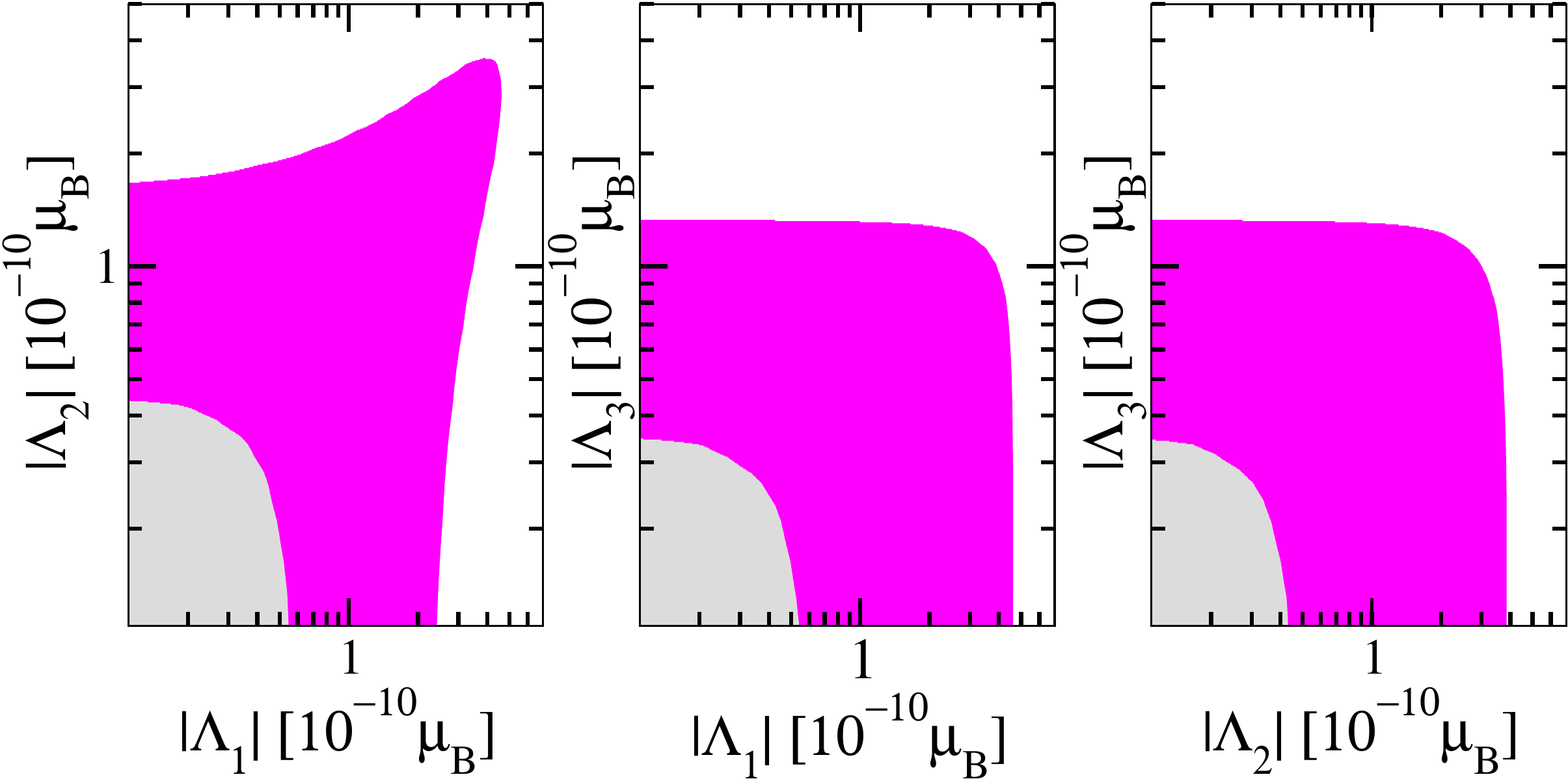}
\caption{90\% C.L. allowed regions for the transition neutrino
  magnetic moments in the mass basis. The result of this plot was
  obtained for the two parameters $|\Lambda_i|$ vs $|\Lambda_j|$
  marginalizing over the third component. We show the result of a
  combined analysis of reactor and accelerator data with all phases
  set to zero except for $\delta = 3\pi/2$ (magenta region). We also
  show the result of the Borexino data analysis only, that is
  phase-independent (grey region). It is visible that Borexino data
  gives a more stringent constraint. See text for details. }
\label{fig:combined}
\end{figure}
%

\section{Conclusions}

In this work we have analyzed the current status of the constraints on
neutrino magnetic moments. We have presented a detailed discussion of
the constraints on the absolute value of the transition magnetic
moments, as well as the role of the CP phases, stressing the
complementarity of different experiments.
Thanks to the low energies observed, below 1 MeV, and its robust
statistics, the Borexino solar experiment plays a very important role
in constraining the electromagnetic neutrino properties.
Indeed, it provides stringent constraints on the absolute magnitude of
the the transition magnetic moments, which we obtain as
\begin{eqnarray}
|\Lambda_1| & \leq 5.6\times 10^{-11}\mu_B \, , \nonumber \\
|\Lambda_2|  & \leq 4.0\times 10^{-11}\mu_B \, ,  \\
|\Lambda_3| & \leq 3.1\times 10^{-11}\mu_B \nonumber \, ,
\end{eqnarray}
However, the incoherent nature of the solar neutrino flux makes
Borexino insensitive to the Majorana phases which characterize the
transition moments matrix.
Although less sensitive to the absolute value of the transition
magnetic moment strengths, reactor and accelerator experiments provide
the only chance to obtain a hint of the complex CP phases. We
illustrate this fact by presenting the constraints resulting from our
global analysis for different values of the relevant CP phases.
Although less stringent than astrophysical limits say, from globular
clusters~\cite{Raffelt:1990pj,Arceo-Diaz:2015pva} or searches for
anti-neutrinos from the sun~\cite{Miranda:2004nz,Miranda:2003yh},
laboratory limits are model independent and should be further pursued.
Indeed, as we have illustrated, improved reactor and accelerator
neutrino experiments will be crucial towards obtaining the detailed
structure of the neutrino electromagnetic properties.

\section{Addendum}

After the publication of this work we noticed that the uncertainties
in the considered backgrounds in Borexino may affect our reported
limit on the neutrino magnetic moment from Borexino data.
Indeed, we have found that a more precise treatment of the
uncertainties in the total normalization of these backgrounds results
in a weaker sensitivity on the neutrino magnetic moment.
This point will be hopefully improved in the near future thanks to the
purification processes carried out in the second phase of the Borexino
experiment.
Meanwhile, however, we think it would be more reliable to adopt the
bound on the neutrino magnetic moment reported by Borexino: $\mu_\nu <
5.4\times 10^{-11}\mu_B$~\cite{Arpesella:2008mt}.
In this case, our Fig.~\ref{fig:combined} should be replaced by the
new version, Fig.~\ref{fig:combined-new}. There, we have added a new
region obtained by allowing the free normalization of backgrounds in
Borexino. The grey region, in contrast, has been obtained for fixed
normalization of the backgrounds in Borexino.
We thank Gianpaolo Bellini from the Borexino Collaboration for
pointing out this issue.

\begin{figure}[t]
\centering
\includegraphics[width=0.75\linewidth]{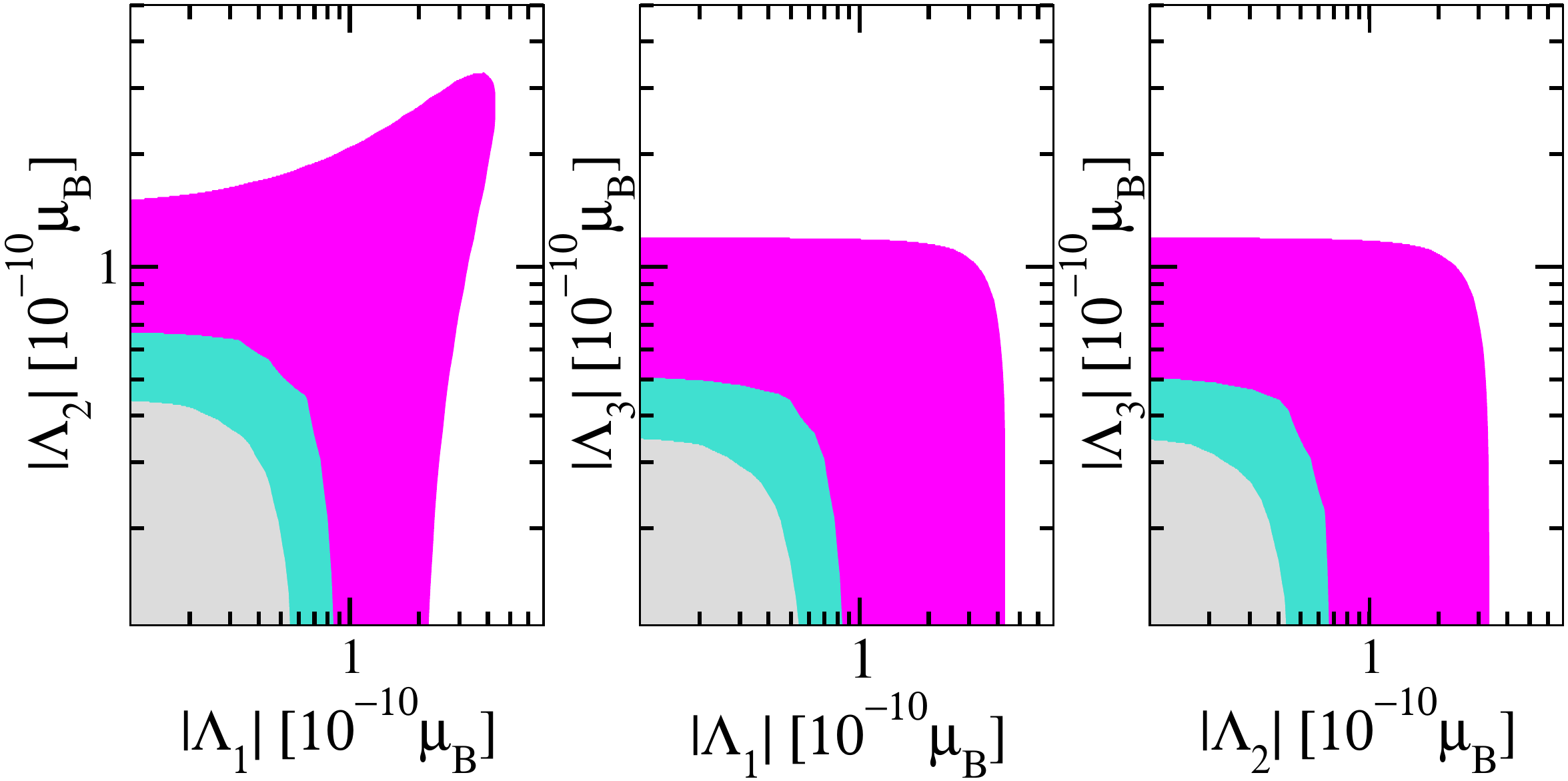}
\caption{90\% C.L. allowed regions for the transition neutrino
  magnetic moments in the mass basis. The result of this plot was
  obtained for the two parameters $|\Lambda_i|$ vs $|\Lambda_j|$
  marginalizing over the third component. We show the result of a
  combined analysis of reactor and accelerator data with all phases
  set to zero except for $\delta = 3\pi/2$ (magenta region). The
  phase-independent results from Borexino are shown in 
  grey (turquoise) for fixed (free) normalization backgrounds in the Borexino data analysis.}

\label{fig:combined-new}
\end{figure}
%

\section*{Acknowledgements}

Work supported by MINECO grants FPA2014-58183-P, Multidark CSD2009-
00064 and SEV-2014-0398 (MINECO); by EPLANET, by the CONACyT grant
166639 (Mexico) and the PROMETEOII/2014/084 grant from Generalitat
Valenciana.
M.~T\'ortola is also supported by a Ramon y Cajal contract of the
Spanish MINECO.
The work of A.Parada was partially supported by Universidad Santiago
de Cali under grant 935-621114-031.

\bibliography{mm-phases}
\end{document}